\pdfoutput=1

\documentclass[11pt]{article}

\usepackage{authblk}
\usepackage{EMNLP2022}

\usepackage{times}
\usepackage{latexsym}

\usepackage[T1]{fontenc}

\usepackage[utf8]{inputenc}

\usepackage{microtype}

\usepackage{inconsolata}

\usepackage{graphicx}
\usepackage{amsmath}
\usepackage{amssymb}
\usepackage{booktabs}
\usepackage{epsfig}
\usepackage{microtype}
\usepackage{subcaption}
\usepackage{threeparttable}
\usepackage[utf8]{inputenc} 
\usepackage{url}            
\usepackage{amsfonts}       
\usepackage{nicefrac}       
\usepackage{xcolor}
\usepackage{layouts}
\usepackage{float}
\usepackage[frozencache,cachedir=.]{minted}        
\usepackage{listings}

\newfloat{code}{htp}{lop}
\floatname{code}{Figure}

\usepackage{amssymb,amsopn}
\usepackage{bm} 
\usepackage{bbm}
\usepackage{enumerate}

\usepackage[capitalize]{cleveref}
\crefname{section}{Sec.}{Secs.}
\Crefname{section}{Section}{Sections}
\Crefname{table}{Table}{Tables}
\crefname{table}{Tab.}{Tabs.}

\newcommand{\vct}[1]{\boldsymbol{\mathbf{#1}}} 

\title{Normalized Contrastive Learning for Text-Video Retrieval}

\author[1]{Yookoon Park\thanks{\hspace{3pt} Work done during internship at Meta} \hspace{1pt}}

\affil[1]{Columbia University, New York, 10027, USA}
\affil[ ]{\texttt{yookoon.park@columbia.edu}}

\author[2]{Mahmoud Azab}\author[2]{Bo Xiong}
\author[2]{Seungwhan Moon} 
\author[2]{\authorcr Florian Metze}
\author[2]{Gourab Kundu}
\author[2]{Kirmani Ahmed}
\affil[2]{Meta, Menlo Park CA 94025, USA}
\affil[ ]{\texttt{\{azab,bxiong,shanemoon,fmetze,gkfacebookny,kirmani\}@fb.com}}

\begin{document}
\maketitle
\begin{abstract}
Cross-modal contrastive learning has led the recent advances in multimodal retrieval with its simplicity and effectiveness. 
In this work, however, we reveal that cross-modal contrastive learning suffers from incorrect normalization of the sum retrieval probabilities of each text or video instance. Specifically, we show that many test instances are either over- or under-represented during retrieval, significantly hurting the retrieval performance. To address this problem, we propose Normalized Contrastive Learning (NCL) which utilizes the Sinkhorn-Knopp algorithm to compute the instance-wise biases that properly normalize the sum retrieval probabilities of each instance so that every text and video instance is fairly represented during cross-modal retrieval.
Empirical study shows that NCL brings consistent and significant gains in text-video retrieval on different model architectures, with new state-of-the-art multimodal retrieval metrics on the ActivityNet, MSVD, and MSR-VTT datasets without any architecture engineering. 
\end{abstract}

\section{Introduction}
With the advent of large-scale multimodal data and transformer-based architectures, cross-modal contrastive learning has contributed to the recent advances in multimodal representation learning \cite{luo2021clip4clip, radford2021learning}. Cross-modal contrastive learning provides a simple, yet highly effective approach for learning representations from multimodal data without supervisions. In particular, CLIP (Contrastive Language-Image Pretraining) \cite{radford2021learning} learns image-text representations using vision and text transformer encoders on millions of image-text data from the Web and demonstrates that the learned vision and text encoders can perform zero-shot transfer on various vision tasks. More recently, CLIP4Clip \cite{luo2021clip4clip} extends the pretrained CLIP model and finetunes it on text-video datasets for embedding-based text-video retrieval, achieving state-of-the-art performance. 


In embedding-based retrieval, the retrieval probabilities of each text or video instance are defined by their embedding similarity to their cross-modal queries. 
When there is one-to-one correspondence between text and video instances as in many standard benchmarks, one would expect that the retrieval probabilities of a video summed over the text queries should be normalized to 1 and vice versa, so that in overall, all text and video instances are equally represented during retrieval. 
However, we show that in practice, cross-modal contrastive learning suffers from significant normalization errors of the sum retrieval probabilities of each instance (\cref{fig:normalization_error}). 
This suggests that many test instances are either over- or under-represented during retrieval, leading to high false positive and false negative rates (\cref{fig:false_rates}) and consequently harming the text-video retrieval performance.

\begin{figure*}[t]
\centering
\begin{subfigure}[b]{0.35\textwidth}
    \centering
    \includegraphics[width=\textwidth]{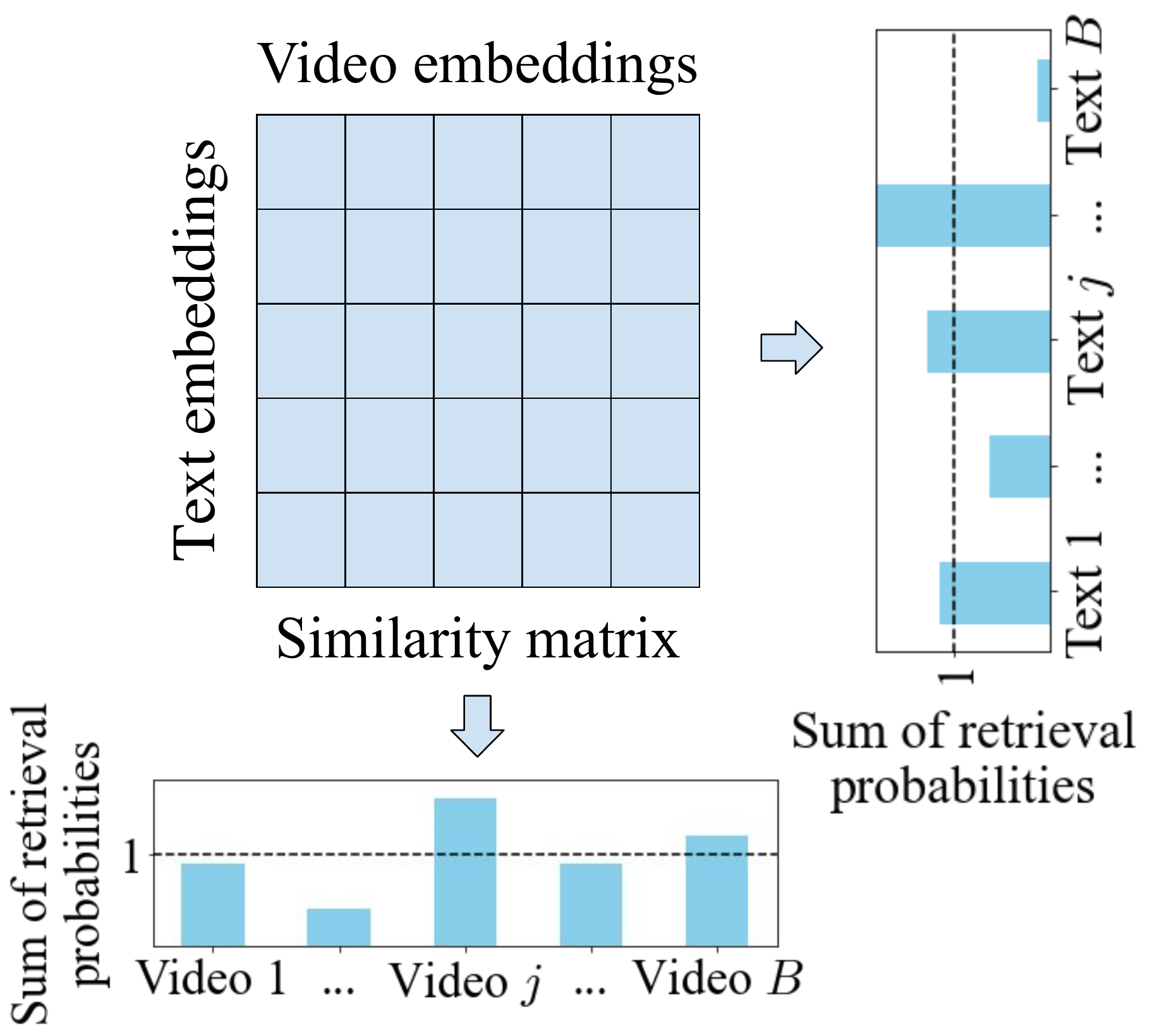}
    \caption{Contrastive learning}
\end{subfigure}
\hspace{0.02\textwidth}
\begin{subfigure}[b]{0.37165\textwidth}
    \centering
    \includegraphics[width=\textwidth]{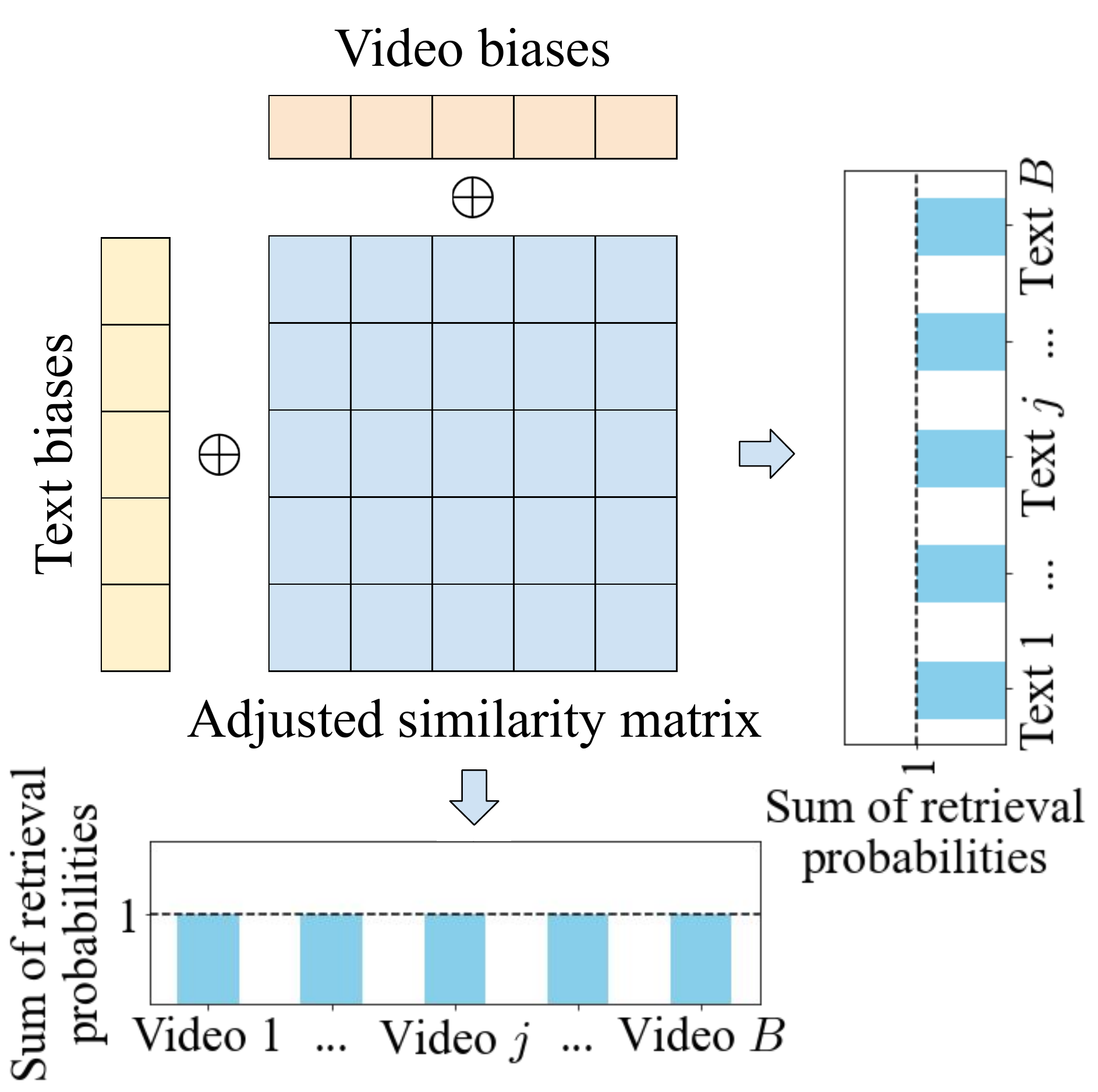}
    \caption{Normalized contrastive learning}
\end{subfigure}
\caption{Overview of the proposed approach. (a) Cross-modal contrastive learning suffers from incorrect normalization of the sum retrieval probabilities of each text/video instance (b) Normalized contrastive learning computes the instance-wise biases that normalize the sum retrieval probabilities so that all instances are fairly represented}
\label{fig:similarity_matrix}
\end{figure*}

To address this problem, we propose \textit{Normalized Contrastive Learning (NCL)} which computes instance-wise biases using the Sinkhorn-Knopp algorithm \cite{cuturi2013sinkhorn} and adjusts the cross-modal embedding similarity scores so that the sum retrieval probabilities of each instance are properly normalized to 1 (\cref{fig:similarity_matrix}).
At test time where the test queries are not known a priori, we show that we can approximate the test query distribution by storing a subset of train queries during training in a queue. We show that this approach consistently reduces the normalization errors (\cref{fig:normalization_error}) and thereby improves the text-video retrieval performance significantly. 

We evaluate NCL on text-video retrieval on the ActivityNet, MSVD, and MSR-VTT datasets. Empirical results show that NCL consistently improves both text-to-video and video-to-text retrieval across all datasets on different base architectures, including those of CLIP4Clip \cite{luo2021clip4clip} and SSB \cite{patrick2021support} and further advances state-of-the-art text-video retrieval performance without any architecture engineering. 

In summary, our contributions are:
\begin{enumerate}
    \item Revealing that cross-modal contrastive learning suffers from incorrect normalization of the sum retrieval probabilities of each instance and how its text-video retrieval performance is impaired by this problem (\cref{sec:normalization_error}). 
    \item Proposing the novel approach of Normalized Contrastive Learning (NCL) to address the normalization errors in cross-modal contrastive learning.
    NCL computes instance-wise biases using the Sinkhorn-Knopp algorithm \cite{cuturi2013sinkhorn} to adjust the cross-modal similarity scores so that every instance is fairly represented during retrieval (\crefrange{sec:normalized_contrastive_learning}{sec:inference}).
    \item Establishing new state-of-the-art results on text-video retrieval tasks on multiple benchmark datasets (ActivityNet, MSVD and MSR-VTT) without any architecture engineering. Moreover, NCL brings consistent and significant gains across different base model architectures including CLIP4Clip \cite{luo2021clip4clip} and SSB \cite{patrick2021support} (\cref{sec:experiments}).
\end{enumerate}

\section{Related Work}
\paragraph{Contrastive representation learning} InfoNCE (Noise Contrastive Estimation of mutual Information) has been developed for learning unsupervised representations of natural images \cite{wu2018unsupervised, oord2018representation, tian2019contrastive, chen2020simple}. It has quickly gained popularity due to its simplicity and effectiveness, leading the state-of-the-art advances in visual representation learning. 
Specifically, it first samples two different views of an image using random data augmentations. 
The views of a same image constitute a positive pair, while negative pairs consist of views of different images in a mini-batch. The encoder is then trained to minimize the cross-entropy loss to classify the positive pairs from the set of negative pairs based on the embedding similarity between the representations. 


Recently, the scope of contrastive learning has been extended to multimodal data \cite{zhang2020contrastive, miech2020end, radford2021learning, amrani2021noise, luo2021clip4clip}. In cross-modal contrastive learning, different modalities of the data are mapped to a shared embedding space using separate encoders for each modality, where their cross-modal similarity is defined. 
In particular, CLIP \citep{radford2021learning} trains its image and text encoders on millions of image-text pairs using cross-modal contrastive learning and demonstrates that the learned representations can perform zero-shot transfer on various vision tasks. 
\citet{li2021align,li2022blip} further improves the vision-language contrastive learning by incorporating additional language modeling and image-text matching losses.

Aside from InfoNCE where each example is assigned its own class label, a recent line of works \cite{asano2020self, asano2020labelling, caron2020unsupervised} assume that the data can be clustered into $K$ latent classes. 
They generate pseudo labels for the data by solving an optimal transport problem with entropy regularization, using the Sinkhorn-Knopp algorithm \cite{cuturi2013sinkhorn}. The encoder is then trained to predict the pseudo-labels of the data.
In contrast to previous work that used the Sinkhorn-Knopp algorithm for self-labeling, in this work we use the algorithm to compute the instance-wise biases that properly normalizes the sum retrieval probabilities of each text or video instance so that all instances are fairly represented during retrieval. 

\paragraph{Embedding-based text-video retrieval} maps video and text data into a shared multimodal embedding space where the similarity between a text-video pair is defined as the cosine similarity of their embeddings. Many recent works build upon pretrained vision and text encoders \cite{liu2019use,miech2019howto100m,miech2020end,liu2021hit,amrani2021noise,patrick2021support,croitoru2021teach,luo2021clip4clip} and finetune their models on downstream text-video datasets. 
For example, \citet{liu2019use} aggregates multiple expert features including objects, motion, appearance, and audio using a collaborative gating mechanism and minimizes ranking loss to align video and text embeddings for retrieval. Similarly, \citet{gabeur2020multi} applies self-attention to video expert features to get video-level representations.
On the other hand, \citet{croitoru2021teach} trains multiple teacher models using different pretrained text encoders and distills their knowledge to a student model. 
\citet{patrick2021support} introduces an auxiliary task of reconstructing the  caption of a video from other similar videos in a mini-batch to improve the learning of multimodal representation. 
\citet{bain2021frozen} tailors vision transformer architecture to train the model on both image and video together and applies curriculum learning by gradually increasing the number of frames the vision encoder takes.
\citet{liu2021hit} builds hierarchical transformers and performs hierarchical cross-modal contrastive matching at both the feature and semantic levels. 
In particular, CLIP4Clip \cite{luo2021clip4clip} achieves state-of-the-art performance on text-video retrieval by loading the pretrained CLIP model \cite{radford2021learning} and finetuning it on text-video datasets using cross-modal contrastive learning. 
However, we show that CLIP4Clip significantly suffers from incorrect normalization of the sum retrieval probabilities of each instance and propose Normalized Contrastive Learning to address this problem. 

\section{Approach}
\label{sec:approach}

\subsection{Cross-modal Contrastive Learning}
\label{sec:contrastive_learning}
We start with a brief introduction of cross-modal contrastive learning.
Given a batch of $B$ ground-truth text-video pairs, video and text encoders map the input to embedding vector pairs $\{(t_1, v_1), ..., (t_B, v_B)\}$ where each embedding lies on the unit hypersphere $S^D$. 
Cross-modal similarity between a text-video pair is defined as the cosine similarity of their embeddings. As the embeddings have the unit norm, the cosine similarity is simply equivalent to their inner product: 
\begin{align}
    \text{sim}(t_i, v_j) = \langle t_i,  v_j \rangle
\end{align}

Text-to-video ($t2v$) and video-to-text ($v2t$) retrieval distributions are defined based on the cross-modal similarity as
\begin{align}
    P_{t2v}(v_j | t_i) &= \frac{\exp(\text{sim}(t_i, v_j))^{1/\gamma}}{\sum_{k=1}^B \exp(\text{sim}(t_i, v_k))^{1/\gamma}} \\
    P_{v2t}(t_j | v_i) &= \frac{\exp(\text{sim}(t_j, v_i))^{1/\gamma}}{\sum_{k=1}^B \exp(\text{sim}(t_k, v_i))^{1/\gamma}},
\end{align}
where $\gamma$ is a temperature parameter that controls the concentration of the distributions. In text-to-video retrieval, text embedding $t_i$ serves as a query for video embedding $v_i$ and vice versa.

Cross-modal contrastive learning \cite{radford2021learning, luo2021clip4clip} minimizes the sum of text-to-video and video-to-text cross-entropy losses:
\begin{align}
    \mathcal{L}_{t2v} 
    &= -\frac{1}{B} \sum_{i=1}^B \log \frac{\exp(\text{sim}(t_i, v_i))^{1/\gamma}}{\sum_{k=1}^B \exp(\text{sim}(t_i, v_k))^{1/\gamma}} \nonumber \\
    \mathcal{L}_{v2t} 
    &= -\frac{1}{B} \sum_{i=1}^B \log \frac{\exp(\text{sim}(t_i, v_i))^{1/\gamma}}{\sum_{k=1}^B \exp(\text{sim}(t_k, v_i))^{1/\gamma}} \nonumber \\
    \mathcal{L} &= \frac{1}{2} (\mathcal{L}_{t2v} + \mathcal{L}_{v2t}) 
    \label{eq:contrastive_loss}
\end{align}
In particular, CLIP \cite{radford2021learning} is trained on millions of image-text pairs using cross-modal contrastive learning. 
CLIP4Clip \cite{luo2021clip4clip} finetunes the pretrained CLIP model for text-video retrieval, achieving state-of-the-art performance. 

\subsection{Normalization Errors in Retrieval}
\label{sec:normalization_error}
\begin{figure}[t]
\centering
\begin{subfigure}[b]{0.24\textwidth}
    \centering
    \includegraphics[width=\textwidth]{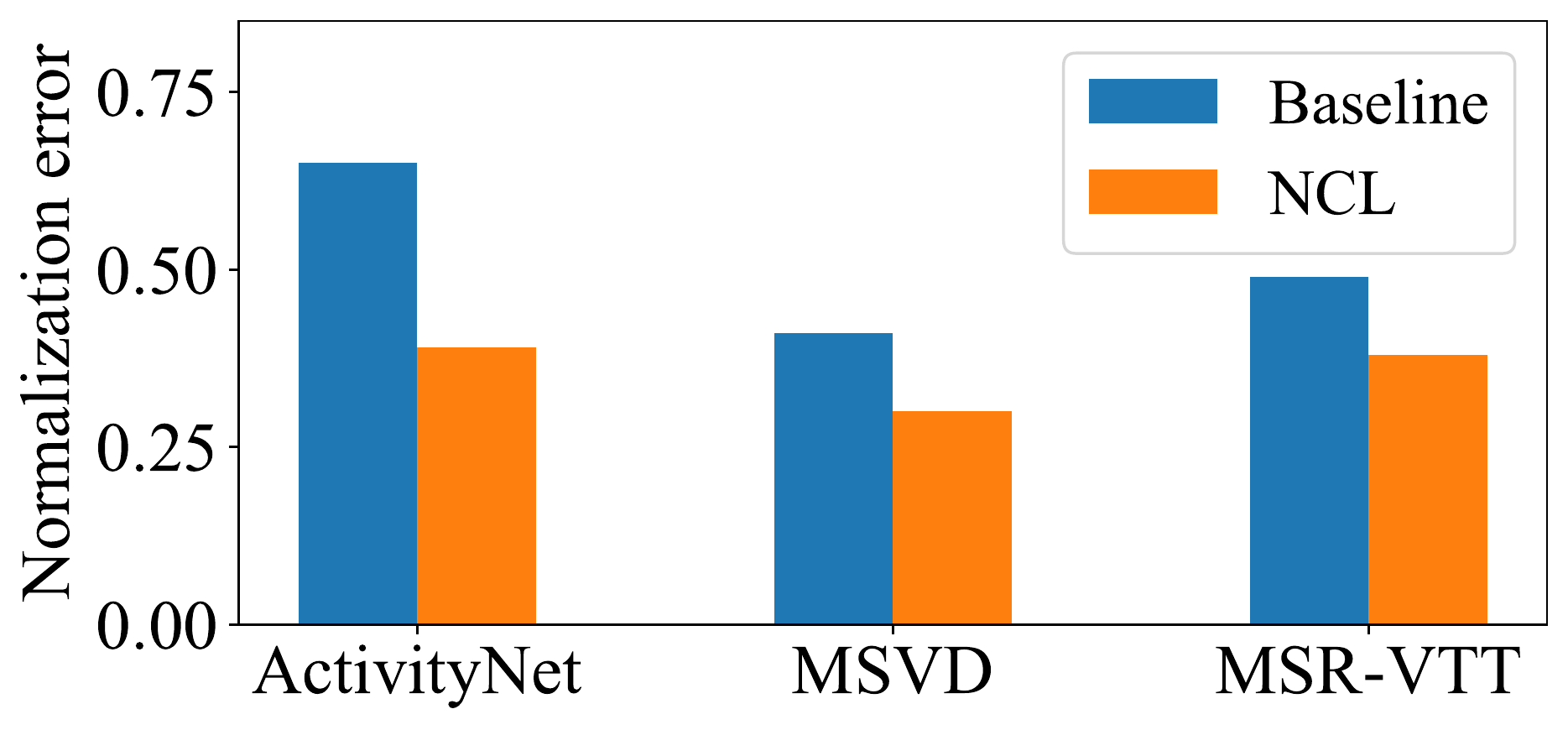}
    \caption{Text-to-video}
\end{subfigure}
\hspace{-0.01\textwidth}
\begin{subfigure}[b]{0.24\textwidth}
    \centering
    \includegraphics[width=\textwidth]{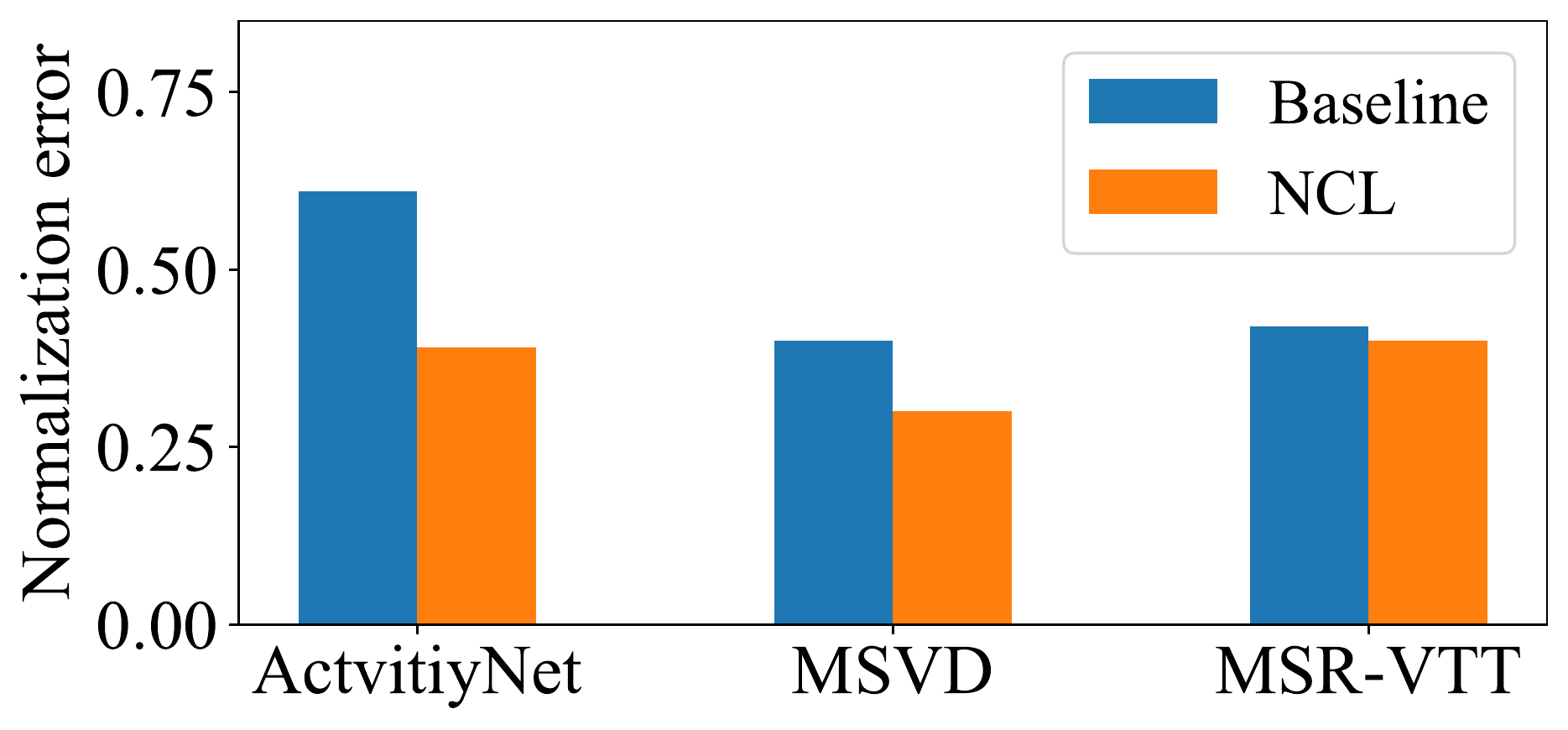}
    \caption{Video-to-text}
\end{subfigure}
\caption{Normalization errors (e.g. \cref{eq:t2v_normalization_error}) of the sum retrieval probabilities at test time. The baseline is CLIP4Clip \cite{luo2021clip4clip} using cross-modal contrastive learning. Normalized Contrastive Learning (NCL) consistently reduces the normalization errors across the datasets. The remaining errors of NCL comes from approximating the unknown test query distribution with a subset of train queries.}
\label{fig:normalization_error}
\end{figure}

When there is one-to-one correspondence between the video and text instances as in many standard benchmarks, one would expect that the retrieval probabilities for a video $v_j$ summed over the text queries to be 1, i.e., $\sum_i P_{t2v}(v_j | t_i) = 1, \forall j$ and vice versa: $\sum_i P_{v2t}(t_j | v_i) = 1, \forall j$, so that all instances are equally represented during retrieval. 

However, \cref{fig:normalization_error} shows that  in practice, CLIP4Clip \cite{luo2021clip4clip} trained using cross-modal contrastive learning suffers from significant normalization errors of the sum retrieval probabilities, where we define the text-to-video normalization error as average the absolute deviation of the sum of video retrieval probabilities from 1:
\begin{align}
\mathbb{E}[\, |1 - \sum_{i=1}^N P_{t2v}(v_j|t_i)|\, ], 
\label{eq:t2v_normalization_error}
\end{align}
where $N$ is the number of test text queries. The video-to-text normalization error is defined in a symmetrical manner. 

\begin{figure}[t]
\centering 
\begin{subfigure}[b]{0.24\textwidth}
    \centering
    \includegraphics[width=\textwidth]{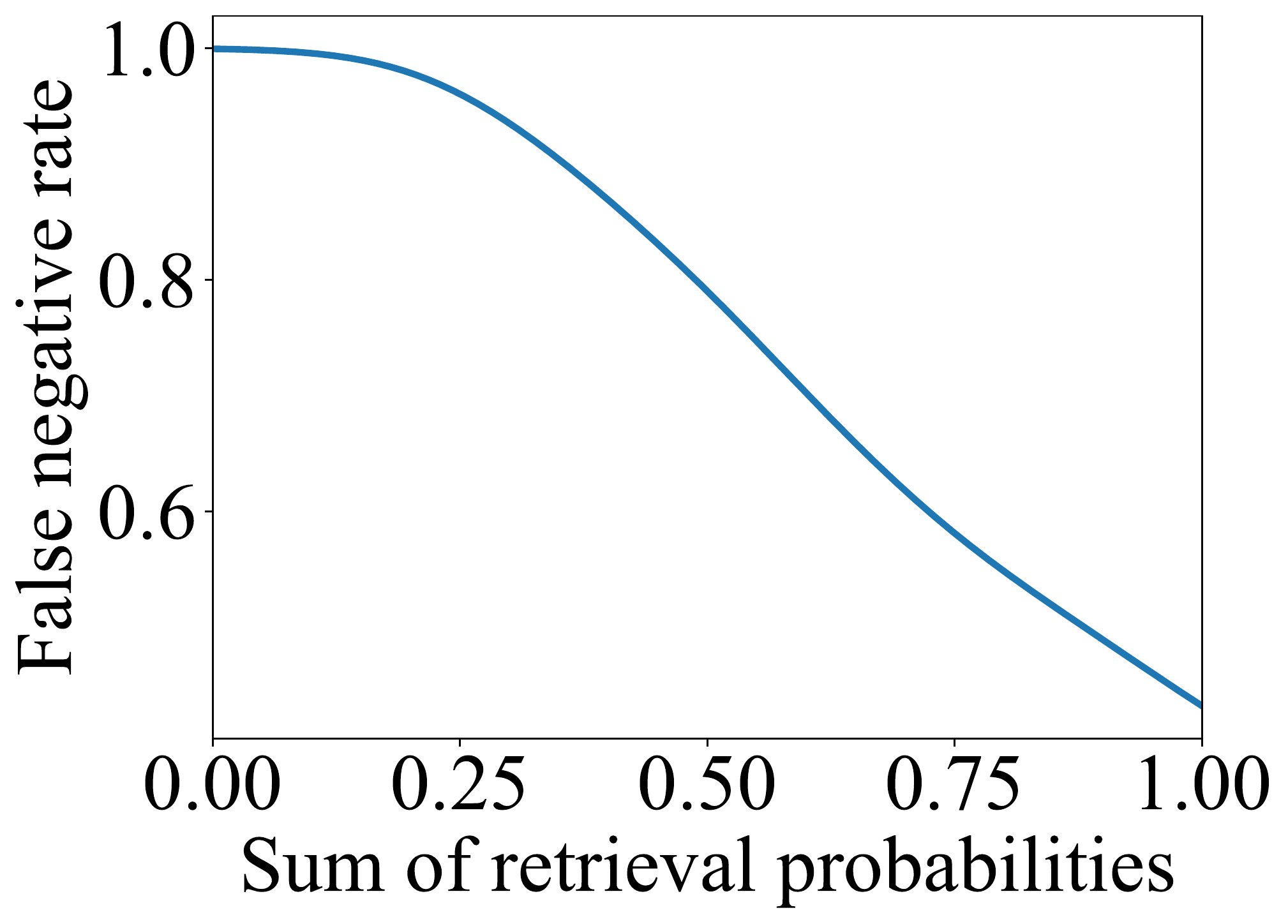}
    \caption{False negative rate}
    \label{fig:false_negative_rate}
\end{subfigure}
\begin{subfigure}[b]{0.223\textwidth}
    \centering
    \includegraphics[width=\textwidth]{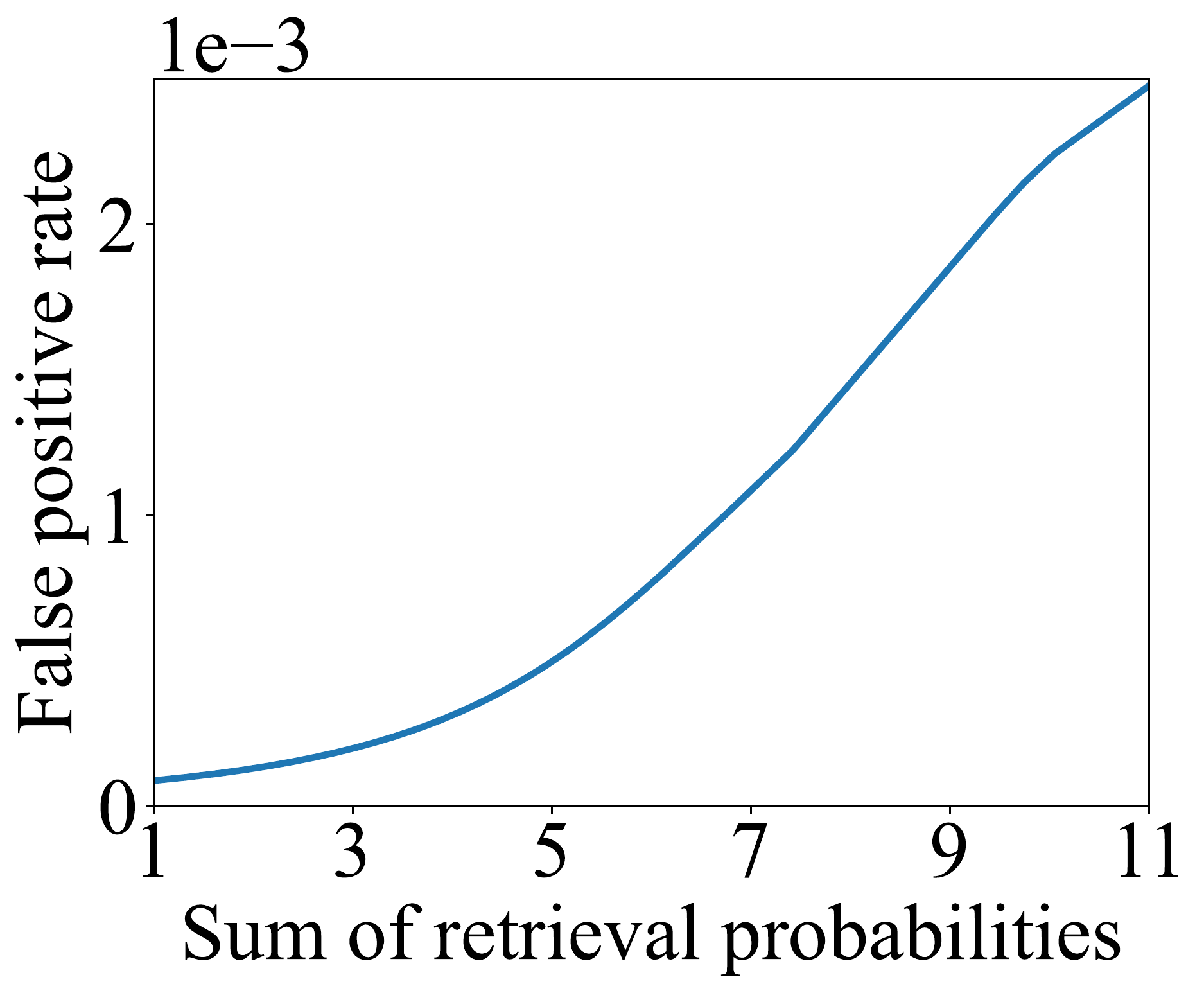}
    \caption{False positive rate}
    \label{fig:false_positive_rate}
\end{subfigure}
\caption{False negative/positive rates vs. sum of retrieval probabilities $\sum_i P_{t2v}(v_j|t_i)$ for a video in text-to-video retrieval on ActivityNet. The false negative/positive rates rapidly increase as the sum of retrieval probabilities of a video deviates from 1}
\label{fig:false_rates}
\end{figure}

Incorrect normalization of the sum retrieval probabilities compromises the retrieval performance:
\begin{enumerate}
    \item If $\sum_{i=1}^N P_{t2v}(v_j | t_i) < 1$, $v_j$ is \textit{under-represented} than the average. It will have higher chance of not being retrieved by the true query $t_j$ (false negative)
    \item If $\sum_{i=1}^N P_{t2v}(v_j | t_i) > 1$, $v_j$ is \textit{over-represented} than the average. It will have higher chance of being wrongly retrieved by irrelevant queries $t_k$ for $k \neq j$ (false positive).
\end{enumerate}
\Cref{fig:false_rates} illustrates these phenomena on ActivityNet, demonstrating how normalization error is correlated with false negative and false positive rates in retrieval. 
Given that cross-modal contrastive learning suffers from significant normalization errors (\cref{fig:normalization_error}), this suggests that its retrieval performance is substantially impaired by incorrect normalization of the sum retrieval probabilities.


\subsection{Normalized Contrastive Learning}
\label{sec:normalized_contrastive_learning}
To address the problem, we propose Normalized Contrastive Learning (NCL) that normalizes the sum retrieval probabilities of each instance so that all instances are equally represented during retrieval. 
First, we introduce instance-wise biases and define the adjusted cross-modal similarity score as
\begin{align}
    \text{sim}(t_i, v_j) = a_i + b_j + \langle t_i, v_j \rangle,
    \label{eq:adjusted_score}
\end{align}
where $a_i$ is a text bias for text $i$ and $b_j$ is a video bias for video $j$.
These instance-wise biases adjust the overall weights of the instances during retrieval. For example, a positive video bias $b_j$ will increase the overall retrieval probabilities for $v_j$, while a negative video bias will decrease the overall retrieval probabilities for $v_j$. 
Therefore, by setting instance-wise biases $a_i, b_j$ to appropriate values, we can properly normalize the retrieval probabilities of text and video instances. 

NCL utilizes the Sinkhorn-Knopp algorithm \cite{cuturi2013sinkhorn} to compute the optimal values of biases for text: $\{a_1^*, \dots, a_B^*\}$, and videos: $\{b_1^*, \dots, b_B^*\}$. 
Specifically, given a non-negative matrix $M \in \mathbb{R}^{m \times n}_{+}$, the Sinkhorn-Knopp algorithm computes the normalization vectors $\alpha \in \mathbb{R}^m_+, \beta \in \mathbb{R}^n_+$ using fixed-point iterations such that 
\begin{align}
    &P = \text{diag}(\alpha) \, M \, \text{diag}(\beta),
\end{align}
is a valid transportation polytope, i.e., 
\begin{align}
    P \vct{1}_n &= \frac{1}{n} \vct{1}_n,  \label{eq:normalized1} \\
    P^T \vct{1}_m &= \frac{1}{m} \vct{1}_m.  \label{eq:normalized2}
\end{align}
In other words, $P$ represents a joint probability distribution for two random variables $X, Y$ such that $P(X=i, Y=j) = p_{ij}$ with uniform marginal constraints $P(X = i) = 1/m$, $P(Y = j) = 1/n$. For retrieval, this means that all instances are equally represented given the set of queries.
In this work, we focus on the standard setting where there is one-to-one correspondence between the video and text captions; hence we assume uniform marginal priors for the instances. 
However, note that the Sinkhorn-Knopp algorithm generalizes to arbitrary prior distributions $P(X), P(Y)$  (e.g. see \citet{asano2020labelling}) and if a video is matched to multiple text captions or vice versa, we can easily modify the priors accordingly so that its sum of retrieval probabilities scales proportionally to the number of matching queries. 
Specifically, the RHS of \crefrange{eq:normalized1}{eq:normalized2} will be placed with simplex vectors $\vct{r}$ and $\vct{c}$ that represent the marginal distributions of text and video instances. The Sinkhorn-Knopp algorithm will normalize the retrieval distributions so that the sum retrieval probabilities of each instance normalizes to its marginal weight.

Given the text-video similarity matrix $S = \{s_{ij}\} \in \mathbb{R}^{B \times B}$, $s_{ij} = \langle t_i, v_j \rangle / \gamma$, we define the non-negative matrix as $M = \exp(S)$. 
After computing the normalization vectors $\alpha, \beta$ for $M$ using the Sinkhorn-Knopp algorithm, the optimal values of the text and video biases are derived as
\begin{align}
    a_i^* &= \gamma \log \frac{\alpha_i}{\sum_k \alpha_k}, \text{ for } i=1, \dots, B, \\
    b_j^* &= \gamma \log \frac{\beta_j}{\sum_k \beta_k}, \text{ for } j=1, \dots, B.
\end{align}
\Cref{fig:sinkhorn_knopp_algorithm} gives a PyTorch-style implementation of the algorithm for computing the instance-wise biases using fixed-point iterations. We set the number of fixed point iterations to 4 and find that the residuals are sufficiently small for our experiments.

\begin{figure}[t]
\small
\setminted{mathescape, escapeinside=||}
\begin{minted}{python}
def sinkhorn_knopp(S, gamma, n_iters):
    M = S.exp()
    beta = 1 / M.sum(0)
    for i in range(n_iters):
        alpha = 1 / (M |@| beta)
        beta = 1 / (alpha |@| M)
    
    alpha /= alpha.sum(keepdims=True)
    beta /= beta.sum(keepdims=True)
    a = gamma * alpha.log()
    b = gamma * beta.log()
    return a, b
\end{minted}
\caption{PyTorch-style implementation of Sinkhorn-Knopp algorithm for computing the optimal biases $a, b$ for similarity matrix $S$. The computed bias vectors are used to adjust the cross-modal similarity scores (\cref{eq:similarity_adjusted})}
\label{fig:sinkhorn_knopp_algorithm}
\end{figure}

NCL uses the computed instance-wise biases to adjust the cross-modal similarity score for retrieval:
\begin{align}
    \text{sim}^*(t_i, v_j) = a_i^* + b_j^* + \langle t_i, v_j \rangle,
    \label{eq:similarity_adjusted}
\end{align}

We can easily verify that the adjusted similarity score of \cref{eq:similarity_adjusted} defines properly normalized retrieval distributions such that 
\begin{align}
\sum_{i=1}^N P^*_{t2v}(v_j|t_i) = \sum_{i=1}^N P^*_{v2t}(t_j|v_i) = 1, \forall ij \label{eq:normalized_distribution}
\end{align}
The proof is in \cref{appendix:normalization_proof}.

During training, NCL computes the optimal biases for the current batch of examples and use the adjusted similarity score (\cref{eq:similarity_adjusted}) for learning.

\subsection{Normalization at Test Time}
\label{sec:inference}
If the test query distribution is known a priori, we can readily construct the test similarity matrix $S = \{s_{ij}\} \in \mathbb{R}^{N \times N}$, $s_{ij} = \langle t_i, v_j \rangle / \gamma$ where $N$ is the number of test instances. In this case, we can directly compute the optimal biases $a^*, b^*$ that normalizes the retrieval weights of the test instances using the algorithm of \cref{fig:sinkhorn_knopp_algorithm}. In this case, the normalization error will be exactly zero. 

In general, however, the test queries may not be known in advance. In such cases, we propose to approximate the unseen test query distribution with a subset of train queries. 
For this purpose, we introduce two query queues that store the last $K$ train queries during training, one for text and video queries each. These query queues can be easily integrated with the existing training loops with negligible computational overhead. The stored queries are only used at test time, as plug-in approximations to the unknown test query distributions.
At test time, we normalize the retrieval probabilities of test instances using the subset of train queries stored in the query queues. 
For text-to-video retrieval, for example, we first construct a pseudo similarity matrix $\widetilde{S} \in \mathbb{R}^{K \times N}$ with $K$ train text queries in the queue and $N$ test video instances. We then apply our normalization algorithm (\cref{fig:sinkhorn_knopp_algorithm}) to compute the video biases $\tilde{b}_1, \dots, \tilde{b}_N$ for the test videos. 
Video-to-text retrieval is handled in a symmetrical manner.
We use the computed biases $\tilde{a}_1, \dots, \tilde{a}_N$ and $\tilde{b}_1, \dots, \tilde{b}_N$ to adjust the cross-modal similarity scores (\cref{eq:adjusted_score}) between the test queries and test instances during retrieval.
Due to the approximation error from using a subset of train queries as a substitute of actual test queries, the normalization error will be nonzero. However, \cref{fig:normalization_error} shows that the proposed approach still consistently reduces the normalization errors on all datasets. 

The computational computational complexity of the test time normalization is $O(KN)$ where $K$ is the size of the train query queue and $N$ is the number of test instances, scaling linearly with the test set size.
Given that embedding-based retrieval already has $O(N^2)$ complexity, NCL does not increase the overall test time complexity as long as $K = O(N)$. 
We study the effects of our approximation and the query queue size on retrieval performance in \cref{sec:effect_of_normalization}.

\section{Experiments}
\label{sec:experiments}

\begin{table*}[t]
\begin{center}
\begin{tabular}{l c c c c c| c c c c c}
\toprule
& \multicolumn{5}{c|}{\textbf{Text $\rightarrow$ Video}} & \multicolumn{5}{c}{\textbf{Video $\rightarrow$ Text}} \\ 
Method  & R@1$\uparrow$ & R@5$\uparrow$ & R@50$\uparrow$ & MdR$\downarrow$ & MnR$\downarrow$ & R@1$\uparrow$ & R@5$\uparrow$ & R@10$\uparrow$ & MdR$\downarrow$ & MnR$\downarrow$\\
\midrule
FSE \cite{zhang2018cross}         & 18.2  & 44.8  & 89.1  & 7   & -     & 16.7 & 43.1 & - & 7 & -    \\ 
CE \cite{liu2019use}              & 18.2  & 47.7  & 91.4  & 6   & 23.1  & 17.7 & 46.6 & - & 6 & 24.4  \\ 
HSE \cite{zhang2018cross}         & 20.5  & 49.3  & -     & -   & -     & 18.7 & 48.1 & -    & - & -  \\ 
TT-CE+ \cite{croitoru2021teach}   & 23.5  & 57.2  & 96.1  & 4   & -     & 23.0 & 56.1 & -    & 4 & -  \\ 
MMT \cite{gabeur2020multi}        & 28.7  & 61.4  & 94.5  & 3.3 & 16.0  & 28.9 & 61.1 & - & 4 & 17.1  \\
SSB \cite{patrick2021support}     & 29.2  & 61.6  & 94.7  & 3   & -     & 28.7 & 60.8 & - & 2 & -  \\ 
CLIP4Clip \cite{luo2021clip4clip} & 40.5  & 72.4 & 98.1  & 2   & 7.4  & 42.5 & 74.1 & 80.6 & 2 & 6.6  \\
\midrule
NCL (Ours) & \textbf{45.9} & \textbf{76.8} & \textbf{98.3} & 2 & \textbf{6.7} & \textbf{46.8} & \textbf{76.5} & \textbf{86.8} & 2 & \textbf{6.2} \\ 
\bottomrule
\end{tabular}
\caption{Multimodal retrieval results on ActivityNet evaluated on the val1 split \cite{caba2015activitynet}. NCL brings significant 13\% and 10\% relative gains in state-of-the-art R@1 on text-to-video and video-to-text retrieval, respectively \\}
\label{table:activitynet}
\vspace{-0.5cm}
\end{center}
\end{table*}

\begin{table*}[t]
\begin{center}
\begin{tabular}{l c c c c c| c c c c c}
\toprule
& \multicolumn{5}{c|}{\textbf{Text $\rightarrow$ Video}} & \multicolumn{5}{c}{\textbf{Video $\rightarrow$ Text}} \\ 
Method  & R@1$\uparrow$ & R@5$\uparrow$ & R@10$\uparrow$ & MdR$\downarrow$ & MnR$\downarrow$ & R@1$\uparrow$ & R@5$\uparrow$ & R@10$\uparrow$ & MdR$\downarrow$ & MnR$\downarrow$\\
\midrule
VSE \cite{kiros2014unifying}            & 12.3  & 30.1  & 42.3  & 14 & - & - & - & - & - & - \\ 
VSE++ \cite{faghri2017vse++}            & 15.4  & 39.6  & 53.0  & 9. & - & - & - & - & - & - \\ 
CE \cite{liu2019use}                    & 19.8  & 49.0  & 63.8  & 6  & - & - & - & - & - & -  \\ 
NoiseE \cite{amrani2021noise}           & 20.3  & 49.0  & 63.3  & 6  & - & - & - & - & - & - \\
TT-CE+ \cite{croitoru2021teach}         & 25.4  & 56.9  & 71.3  & 4  & - & 27.1 & 55.3 & 67.1 & 4 & - \\
SSB \cite{patrick2021support}           & 28.4  & 60.0  & 72.9  & 4  & - & 34.7 & 59.9 & 70.0 & 3 & - \\ 
CLIP4Clip \cite{luo2021clip4clip}       & 46.2	& 76.1	& 84.6	& 2	 & 10.0 & 56.6 & 79.7 & 84.3 & 1 & 7.6 \\
\midrule
NCL (Ours) & \textbf{47.8} & \textbf{77.5} & \textbf{85.9} & 2 & \textbf{9.9} & \textbf{69.6} & \textbf{89.9} & \textbf{95.4} & 1 & \textbf{3.3}\\
\bottomrule
\end{tabular}
\caption{Multimodal retrieval results on MSVD. NCL consistently improves all retrieval metrics with substantial 23\% relative gain in video-to-text R@1 \\}
\label{table:msvd}
\vspace{-0.5cm}
\end{center}
\end{table*}

We evaluate Normalized Contrastive Learning (NCL) on multimodal retrieval on popular text-video datasets including ActivityNet, MSVD, and MSR-VTT and report recall-at-K (R@K) metrics (higher is better), median and mean rank (lower is better). The main goals of the experiments are:
\begin{enumerate}
    \item Compare NCL to state-of-the-art models in text-video retrieval using different model architectures (\cref{sec:comparison_sota})
    \item Analyze the effects of the proposed test time normalization method and the size of the query queue on NCL's text-video retrieval performance (\cref{sec:effect_of_normalization})
\end{enumerate}

For a fair comparison, we assume that the test query distribution is not known in advance and use train query queues for test time normalization. 

\textbf{Datasets}
\textbf{ActivityNet} \cite{krishna2017dense,caba2015activitynet} is a collection of 20,000 YouTube videos. Following \cite{zhang2018cross,gabeur2020multi}, we concatenate the text descriptions of a video into one paragraph and perform video-paragraph retrieval on ActivityNet. We use the 'val1' split for evaluation which contains 5K videos and paragraphs. The train split has 10K video-paragraphs.  
\textbf{MSVD} \cite{chen2011collecting} has 1,970 videos with each video having approximately 40 captions. The train split has 1200 videos, with 100 videos in the validation and 760 videos in the test split. In MSVD, each video in the test set has multiple captions associated with it. 
\textbf{MSR-VTT} \cite{xu2016msr} consists of 10,000 videos with 20 text captions per each video. We use the 9K train split with 180K captions and the 1K test split following \cite{yu2018joint}. The test split only contains one caption per video. 

\begin{table*}[t]
\begin{center}
\begin{tabular}{l c c c c c | c c c c c}
\toprule
& \multicolumn{5}{c|}{\textbf{Text $\rightarrow$ Video}} & \multicolumn{5}{c}{\textbf{Video $\rightarrow$ Text}} \\ 
Models  & R@1$\uparrow$ & R@5$\uparrow$ & R@10$\uparrow$ & MdR$\downarrow$ & MnR$\downarrow$ & R@1$\uparrow$ & R@5$\uparrow$ & R@10$\uparrow$ & MdR$\downarrow$ & MnR$\downarrow$\\
\midrule
JSFusion \cite{yu2018joint}     & 10.2  & 31.2  & 43.2  & 13 & -    & - & - & - & - & - \\ 
HT100M \cite{miech2019howto100m}& 12.1  & 35.0  & 48.0  & 12 & -    & 16.8  & 41.7  & 55.1  & 8 & - \\ 
JPoSE \cite{wray2019fine}       & 14.3  & 38.1  & 53.0  & 9  & -    & 16.4  & 41.3  & 54.4  & 8.7 & - \\ 
CE \cite{liu2019use}            & 20.9  & 48.8  & 62.4  & 6  & -    & 20.6  & 50.3  & 64.0  & 5.3 & - \\ 
MMT \cite{gabeur2020multi}      & 26.6  & 57.1  & 69.6  & 4  & 24.0 & 27.0  & 57.5  & 69.7  & 3.7 & 21.3  \\
TT-CE+ \cite{croitoru2021teach} & 29.6  & 61.6  & 74.2  & 3  & -    & 32.1  & 62.7  & 75.0  & 3 & - \\
SSB \cite{patrick2021support}   & 30.1  & 58.5  & 69.3  & 3  & -    & 28.5  & 58.6  & 71.6  & 3 & - \\
HiT \cite{liu2021hit}           & 30.7  & 60.9  & 73.2  & 2.6& -    & 32.1  & 62.7  & 75.0  & 3 & - \\
CLIP4Clip\cite{luo2021clip4clip}& 43.1  & 70.4  & 80.8  & 2  & 16.2 & 43.1  & 70.5  & \textbf{81.2}  & 2 & \textbf{12.4} \\
\midrule
NCL (Ours) & \textbf{44.9}  & \textbf{71.2}  & \textbf{81.5}  & 2 & \textbf{15.5} & \textbf{44.9}  & \textbf{71.8}  & 80.7  & 2 & 12.8 \\ 
\bottomrule
\end{tabular}
\caption{Multimodal retrieval results on MSR-VTT evaluated on the 1K test split \cite{yu2018joint}. NCL again improves most of the retrieval metrics \\}
\label{table:msrvtt}
\vspace{-0.5cm}
\end{center}
\end{table*}

\textbf{Architecture}
We use the state-of-the-art architecture of CLIP4Clip \cite{luo2021clip4clip} with the code released by the authors. CLIP4Clip is based on the pretrained CLIP model \cite{radford2021learning} which consists of transformer-based vision and text encoders trained on large-scale image-text data. 
For the experiments, we adopt the mean pooling (meanP) architecture for aggregating the frame features as it was shown to deliver the most consistent performance on both text-to-video and video-to-text retrieval across different datasets \cite{luo2021clip4clip}. 
We do not add any trainable model components and use the CLIP4Clip architecture as-is. 

Implementations details are described in \cref{appendix:implementation_details}.



\begin{table*}[t]
\begin{center}
\begin{tabular}{l c c c c c| c c c c c}
\toprule
& \multicolumn{5}{c|}{\textbf{Text $\rightarrow$ Video}} & \multicolumn{5}{c}{\textbf{Video $\rightarrow$ Text}} \\ 
Dataset/Model  & R@1$\uparrow$ & R@5$\uparrow$ & R@50$\uparrow$ & MdR$\downarrow$ & MnR$\downarrow$ & R@1$\uparrow$ & R@5$\uparrow$ & R@10$\uparrow$ & MdR$\downarrow$ & MnR$\downarrow$\\
\midrule
\multicolumn{6}{l|}{\textit{ActivityNet}}\\
SSB \cite{patrick2021support}     & 26.8  & 58.1  & 93.5  & 3   & -     & 25.5 & 57.3 & - & 3 & -  \\ 
SSB + NCL & \textbf{32.8} & \textbf{64.8} & \textbf{96.8} & \textbf{2} & 11.2 & \textbf{33.0} & \textbf{65.0} & \textbf{76.2} & \textbf{2} & 11.4 \\ 
\midrule
\multicolumn{6}{l|}{\textit{MSVD}}\\
SSB \cite{patrick2021support}     & 23.0  & 52.8  & 65.8  & 5   & -     & 27.3 & 50.7 & 60.8 & 5 & -  \\ 
SSB + NCL  & \textbf{24.7} & \textbf{55.3} & \textbf{68.1} & \textbf{3} & 21.6 & \textbf{29.3} & \textbf{54.0} & \textbf{64.0} & \textbf{4} & 45.2 \\ 
\midrule
\multicolumn{6}{l|}{\textit{MSR-VTT}}\\
SSB \cite{patrick2021support}     & 27.4  & 56.3  & 67.7  & 3   & -     & 26.6 & 55.1 & 67.5 & 3 & -  \\ 
SSB + NCL & \textbf{28.5} & \textbf{57.8} & \textbf{70.5} & 3 & 22.3 & \textbf{28.5} & \textbf{57.3} & \textbf{69.4} & 3 & 22.8 \\ 
\bottomrule
\end{tabular}
\caption{Text-video retrieval using the Support Set Bottleneck (SSB) \cite{patrick2021support} architecture on ActivityNet, MSVD and MSR-VTT. We report the results without additional pretraining on the HowTo100M dataset. The results show that NCL brings consistent gains across all datasets regardless of the base architecture \\}
\label{table:ssb}
\vspace{-0.5cm}
\end{center}
\end{table*}

\subsection{Comparison to State of the Art}
\label{sec:comparison_sota}
We compare NCL with the state-of-the-art multimodal retrieval models  \cite{croitoru2021teach,patrick2021support,liu2021hit,luo2021clip4clip} on ActivityNet, MSVD, and MSR-VTT. NCL uses the same architecture as CLIP4Clip \cite{luo2021clip4clip}. 
\Crefrange{table:activitynet}{table:msrvtt} summarize the results. On all datasets, NCL brings significant improvements on state-of-the-art recall metrics in both text-to-video and video-to-text retrieval across different datasets. 
On ActivityNet, NCL gives 13\% and 10\% relative gains on text-to-video and video-to-text R@1, respectively, compared to the previous state-of-the-art of CLIP4Clip \cite{luo2021clip4clip}. 
Especially, the gain on MSVD video-to-text retrieval is substantial with more than 23\% relative boost in R@1 and the mean rank being reduced by more than half. 
This may be attributed to the significant imbalance between the number of video queries and text captions in the MSVD test set with 670 videos and 28K captions, making the proper normalization of the caption retrieval probabilities more crucial.
In addition, NCL improves most of the recall metrics on MSR-VTT. 
We emphasize that these results are achieved without any architecture engineering and the additional computational overhead introduced by NCL is negligible. 

In addition, \cref{table:ssb} presents the results for NCL implemented on the Support-Set Bottleneck (SSB) \cite{patrick2021support} architecture, which demonstrate that NCL consistently improves the retrieval performance regardless of the base architecture. The implementation details for the SSB architecture are described in \cref{appendix:ssb_details}.

\subsection{Analysis of NCL}
\label{sec:effect_of_normalization}

\begin{table*}[t]
\begin{center}
\begin{tabular}{l c c c c c | c c c c c}
\toprule
& \multicolumn{5}{c|}{\textbf{Text $\rightarrow$ Video}} & \multicolumn{5}{c}{\textbf{Video $\rightarrow$ Text}} \\ 
Normalization  & R@1$\uparrow$ & R@5$\uparrow$ & R@10$\uparrow$ & MdR$\downarrow$ & MnR$\downarrow$ & R@1$\uparrow$ & R@5$\uparrow$ & R@10$\uparrow$ & MdR$\downarrow$ & MnR$\downarrow$\\
\midrule
None \cite{luo2021clip4clip} & 40.5  & 72.4. & 83.8  & 2   & 7.4  & 42.5 & 74.1 & 80.6 & 2 & 6.6  \\
Train queries & \textbf{45.9} & \textbf{76.8} & \textbf{86.5} & 2 & \textbf{6.7} & \textbf{46.8} & \textbf{76.5} & \textbf{86.8} & 2 & \textbf{6.2} \\
\midrule
Test queries (oracle) & 54.1 & 80.7 & 89.2 & 1 & 5.7 & 54.2 & 76.7 & 89.4 & 1 & 5.1 \\
\bottomrule
\end{tabular}
\caption{The effect of the proposed approximation (\cref{sec:inference}) on ActivityNet text-video retrieval. The baseline is CLIP4Clip \cite{luo2021clip4clip} without any normalization (first row). 
While using the test queries for normalization gives the best performance (the oracle, last row), using a subset of train queries for normalization still brings significant gains compared to the baseline (second row) even without any knowledge of the test queries \\}
\label{table:activitynet_normalized}
\vspace{-0.5cm}
\end{center}
\end{table*}

\begin{figure}[t]
\centering
\includegraphics[width=0.4\textwidth]{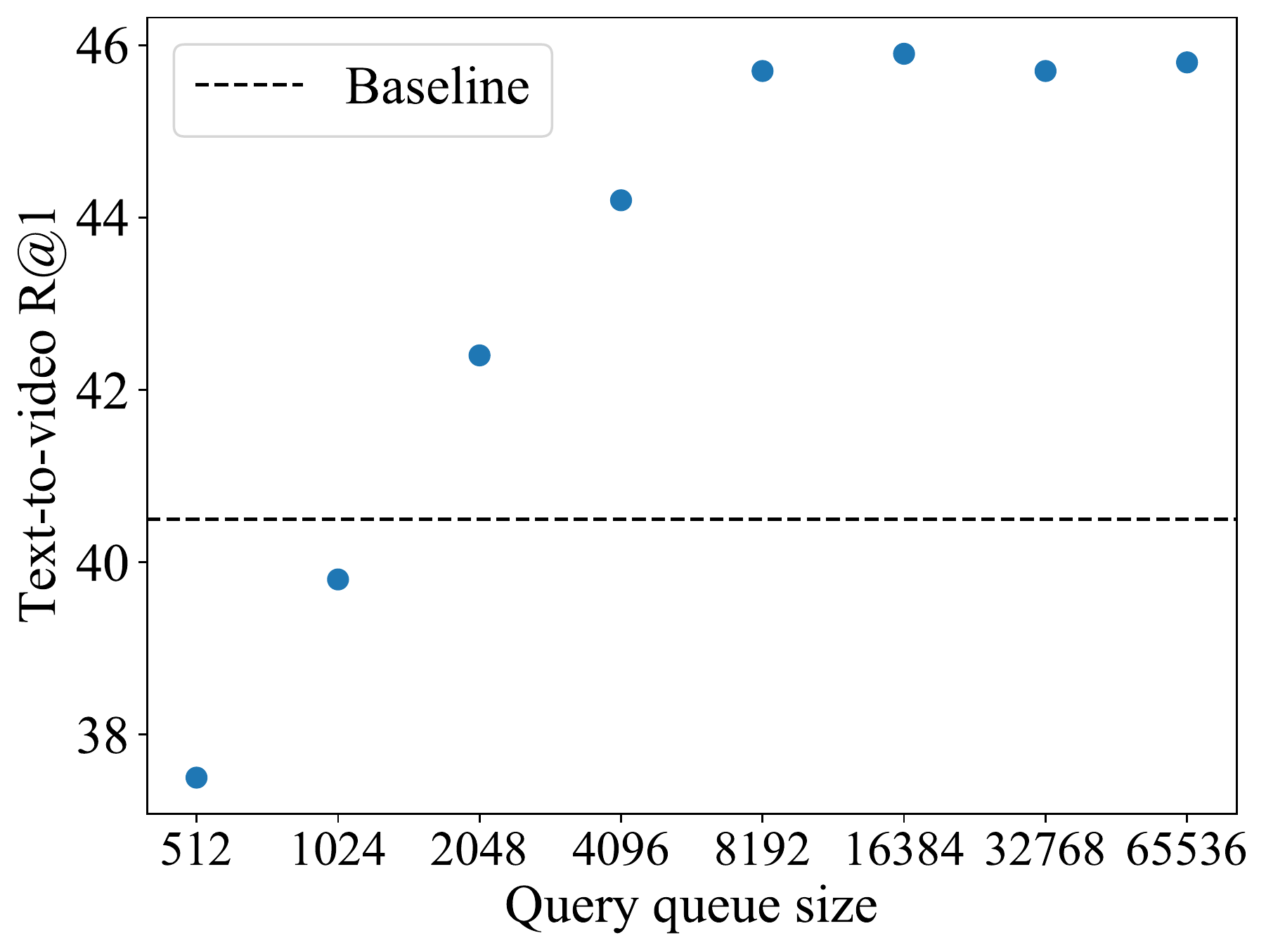}
\caption{Query queue size vs. text-to-video R@1 on ActivityNet. The performance improves as the size of query queue increases and plateaus at about 16K}
\label{fig:memory_queue_size} 
\end{figure}

\begin{figure*}[t]
\centering
\includegraphics[width=1.0\textwidth]{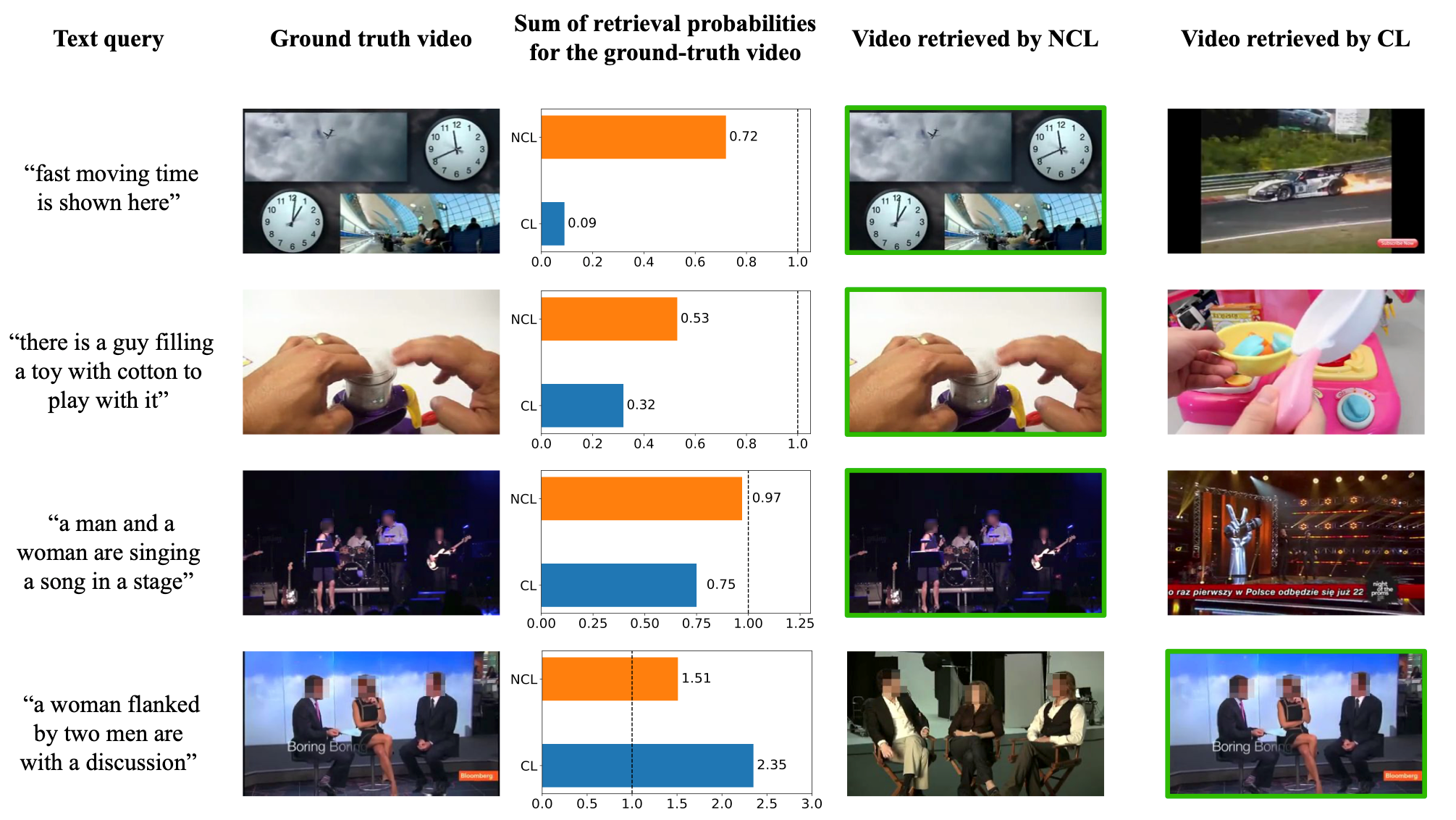}
\caption{Example retrieval results for text-to-video retrieval on MSR-VTT comparing Normalized Contrastive Learning (NCL) to Contrastive Learning (CL). NCL consistently reduces the normalization errors in all examples. The first three rows show the examples where the ground-truth videos are under-represented and CL fails to correctly retrieve the videos, while NCL correctly retrieves the ground truth videos
}
\label{fig:examples}
\end{figure*}

\cref{table:activitynet_normalized} studies the effect of the approximation proposed in \cref{sec:inference}. 
Using the test queries directly for normalization gives the best performance (last row). This is the oracle setting where the normalization errors become zero. However, in general, the test queries may not be known in advance and we approximate the unknown test query distribution using a subset of train queries stored in the query queues. 
We observe that this approximation reduces the test normalization errors (\cref{fig:normalization_error}) and brings significant gains compared to the CLIP4Clip baseline \cite{luo2021clip4clip} even without a prior knowledge of test queries. Still, there is a considerable gap when compared to the oracle. 

\Cref{fig:memory_queue_size} presents the effect of the train query queue size on text-video retrieval performance on ActivityNet. 
The performance improves with growing query queue size and plateaus at about 16K queries. Based on this result, we set the query queue size to 16384 in our experiments. 

We visualize example retrieval results for text-to-video retrieval on MSR-VTT in \cref{fig:examples}. It illustrates how the sum retrieval probabilities are not well-normalized in contrastive learning (CL) with the test videos being severely over- or under-represented during retrieval. NCL consistently alleviates the normalization errors. 

\section{Conclusion}
We've presented Normalized Contrastive Learning (NCL) to improve cross-modal contrastive learning for retrieval. NCL applies the Sinkhorn-Knopp algorithm to normalize the retrieval probabilities of text and video instances so that each instance is fairly represented during retrieval. 
When the test queries are not known a priori, NCL approximates the test query distribution with a subset of train queries stored during training. 
Empirical studies show that NCL consistently reduces the normalization errors and brings significant gains in state-of-the-art text-video retrieval performance without any architecture engineering. Moreover, the gains are consistent over different model architectures. 
For future work, it will be worthwhile to explore if NCL can help retrieval tasks in other domains such as image and text, and whether it can improve general representation learning for downstream tasks such as classification. 

\section*{Limitations}
The proposed approach of Normalized Contrastive Learning (NCL) is broadly applicable to general embedding-based retrieval tasks on unimodal or multimodal domains. However, the scope of the empirical studies in this work was limited to the video-text domain only.
Other limitations include that NCL may not generalize well if the test query distribution is significantly different from the training query distribution or if the test query distribution changes dynamically over time. Such discrepancies may be alleviated by storing the previous test queries on the fly in the query queue to dynamically adapt to the unseen test query distribution. We leave related investigations to future work. 


\section*{Ethics Statement}
The proposed approach retrieves relevant videos given a text query and vice versa. The encoders for video and text were pretrained on large collection of image-text data and finetuned on video-text datasets and may reflect biases in those data, including those with potentially negative societal impacts. In addition, there is a risk that the model will be used for tasks of societal sensitivity, such as surveillance. These concerns call for caution and consideration when deploying or building upon the proposed model.

\bibliography{references}
\bibliographystyle{acl_natbib}

\appendix

\section{Proof for \cref{eq:normalized_distribution}}
\label{appendix:normalization_proof}
We prove that the adjusted cross-modal similarity score (\cref{eq:similarity_adjusted}) using the instance-wise biases computed by the algorithm of \cref{fig:sinkhorn_knopp_algorithm} properly normalizes the retrieval distributions:
\begin{align}
    &\sum_{i=1}^B P_{t2v}^*(v_j | t_i) \\
    &= \sum_{i=1}^B \frac{\exp(\text{sim}^*(t_i, v_j))^{1/\gamma}}{\sum_{k=1}^B \exp(\text{sim}^*(t_i, v_k))^{1/\gamma}} \\
    &= \sum_{i=1}^B \frac{\exp(a_i^* + b_j^* + \langle t_i, v_j \rangle)^{1/\gamma}}{\sum_{k=1}^B \exp(a_i^* + b_k^* + \langle t_i, v_k \rangle)^{1/\gamma}} \\
    &= \sum_{i=1}^B \frac{\alpha_i \beta_j M_{ij}}{\sum_{k=1}^B \alpha_i \beta_k M_{ik}} \\
    &= \sum_{i=1}^B \frac{p_{ij}}{\sum_k p_{ik}} = B \sum_{i=1}^B p_{ij} = B \frac{1}{B} = 1,
    \label{eq:retrieval_distribution_normalized}
\end{align}
where the last line follows from the normalization properties (\crefrange{eq:normalized1}{eq:normalized2}).

\section{Implementation Details for CLIP4Clip}
\label{appendix:implementation_details}
We mostly follow the experimental settings of CLIP4Clip \cite{luo2021clip4clip}. 
We initialize the text and vision encoders with CLIP (ViT-B/32) \cite{radford2021learning} and finetune the model using Adam optimizer \cite{kingma2015adam} and a cosine learning rate schedule without restart \cite{loshchilov2017sgdr} with initial learning of 1e-7. The maximum caption token length is set to 32 for MSVD and MSR-VTT, and 64 for AcitivtyNet. The number of frames sampled is 12 for MSVD and MSR-VTT, and 64 for ActivityNet whose videos are generally much longer in length. 
The only hyperparameters we tweak are: (1) the number of training epochs for each dataset and (2) the size of the query queue. These hyperparameters are selected based on their validation performance.  
We train the model for 3, 5, and 10 epochs for MSVD, MSR-VTT, and ActivityNet, respectively. We apply linear learning rate warm-up over the first 10\% of the training iterations for all datasets except for MSVD where we find that using warm-up slightly degrades the performance.
For test time normalization, NCL introduces two query queues to store a subset of train queries during training, one for text and video queries each. The query queue size is set to 16384. 

\section{Implementation Details for SSB}
\label{appendix:ssb_details}
We describe the experimental details for the experiments using the Support-Set Bottleneck (SSB) architecture. The code for SSB was provided by the authors. 

\textbf{Architecture}.
SSB uses the pretrained T5-base model \cite{raffel2020exploring} for its text encoder. For vision, it first extracts motion and appearance features using the 34-layer R(2+1)-D model \cite{tran2018closer} pretrained on IG65M \cite{ghadiyaram2019large} and ResNet152 \cite{he2016deep} pretrained on ImageNet \cite{deng2009imagenet}, respectively. It then concatenates the motion and appearance features for its visual input. SSB applies CNN and RNN networks on its text and visual features, respectively, followed by transformer pooling layers. For more details, we refer to the original paper \cite{patrick2021support}.

\textbf{Implementation details}.
In our experiments, we use the SSB model \cite{patrick2021support} without additional pretraining on HowTo100M \cite{miech2019howto100m}. 
SSB employs max-margin triplet ranking loss with hard negative mining \cite{faghri2017vse++} and we replace the margin loss with the NCL loss (Eq. 31 to 33 in the paper). The temperature parameter for NCL is set to 0.07 following  \cite{chen2020simple}. NCL uses query queues of size 16384 to store the queries during training. The scale of the NCL loss is multiplied by 15 in order to approximately match the scale of the original margin loss. We apply dropout with ratio 0.1 for MSVD and MSR-VTT, and 0.0 for ActivityNet. 
For the rest of the hyperparameters, we follow the setting used in \cite{patrick2021support}. 

\section{Multimodal Embedding Space of Cross-modal Contrastive Learning}
\label{sec:embedding_space_analysis}
Many recent works \cite{radford2021learning,luo2021clip4clip,xu2021videoclip,miech2020end,bain2021frozen} have demonstrated the promise of cross-modal contrastive learning for multimodal data. However, its behavior and properties in multimodal environments have nor been well-understood until now. In this section, we study the multimodal embedding space of the CLIP4Clip \cite{luo2021clip4clip} model trained using cross-modal contrastive learning.

\begin{figure}[t]
\centering
\includegraphics[width=0.4\textwidth]{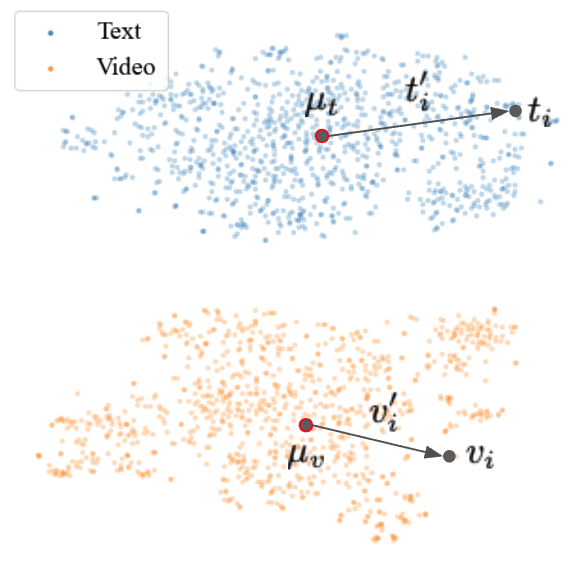}
\vspace{-10pt}
\caption{TSNE visualization of the multimodal embedding space of CLIP4Clip \cite{luo2021clip4clip} on MSR-VTT. The figure perceptually illustrates the modal mean decomposition  (\crefrange{eq:mean_decomposition1}{eq:mean_decomposition2}). $\mu_v, \mu_t$ denote the modal means of video and text embeddings and $v'_i, t'_i$ represents the displacements from their respective modal means. The text and video embeddings are highly clustered to their modalities and do not overlap with each other.}
\label{fig:tsne}
\end{figure}

\begin{figure}[t]
\centering
\includegraphics[width=0.4\textwidth]{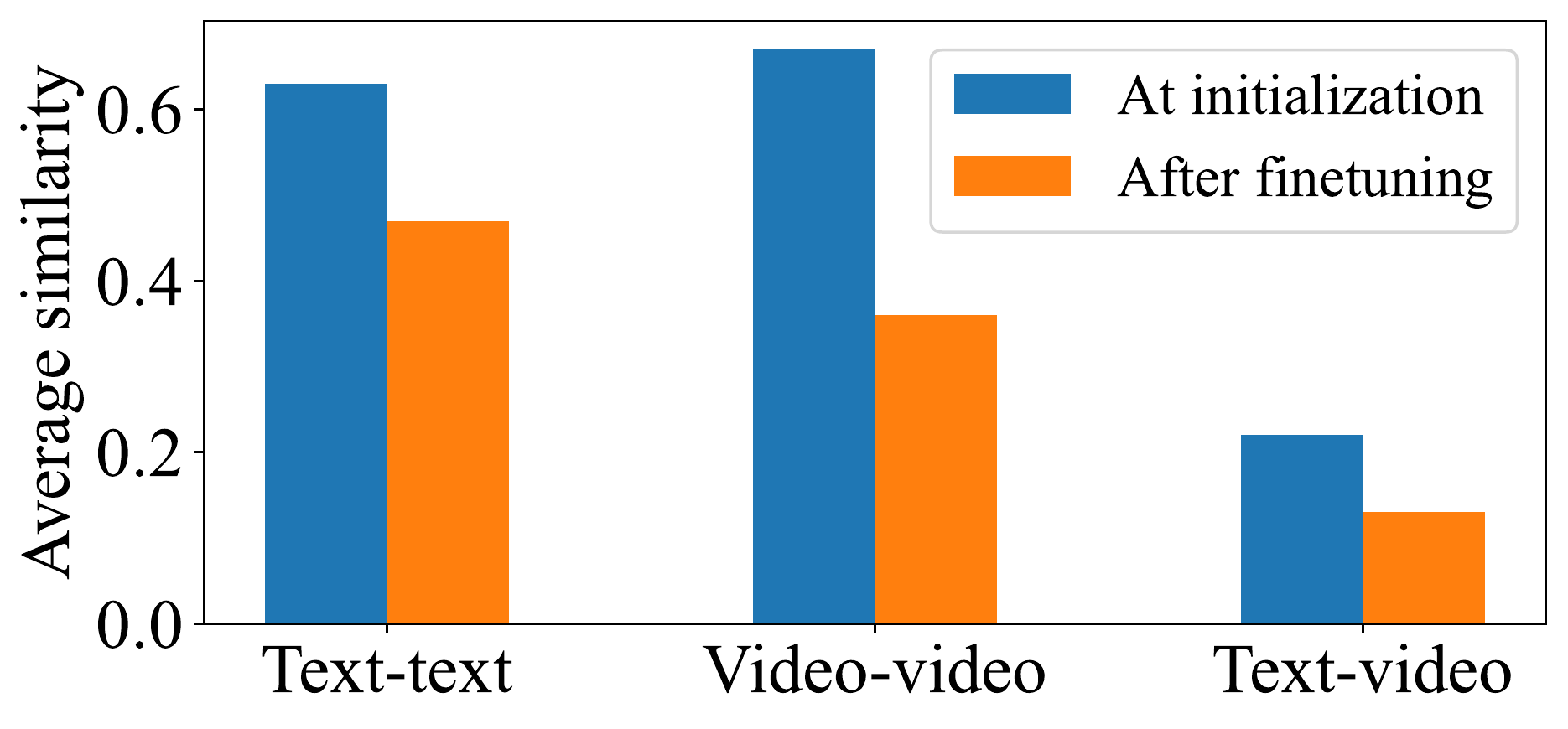}
\vspace{-5pt}
\caption{Average embedding similarities between modalities on MSR-VTT at initialization and after finetuning. 
Even after finetuning, with-in modal similarities are high, showing that the embeddings are clustered to their modalities. On the other hand, relatively low text-video similarity suggests that the embeddings of different modalities do not overlap with each other as depicted in \cref{fig:tsne}}
\label{fig:similarities}
\end{figure}


\Cref{fig:tsne} visualizes the video-text embedding space of the CLIP4CLIP model at initialization on the MSR-VTT dataset. The video and text embeddings do not overlap with each other, being highly clustered to their modalities in the embedding space. 
This is surprising, as the success of contrastive learning in the unimodal setting of natural images has previously been attributed to the alignment and uniformity of the embeddings \cite{wang2020understanding}. The figure shows that the video-text embeddings are neither well-aligned with each other nor evenly distributed on the unit hypersphere.
This can also be confirmed from the average embedding similarities between the modalities in \cref{fig:similarities}, which shows that within-modal embedding similarities are still high even after finetuning on MSR-VTT. In addition, the relatively low average similarity between text and video suggests that the embeddings of different modalities do not overlap as depicted in \cref{fig:tsne}.
This problem may be due to the distribution shift caused by finetuning the pretrained CLIP model on MSR-VTT. While finetuning the model longer might alleviate this issue, we find that longer finetuning harms the retrieval performance due to overfitting. 

To analyze the implications of this phenomenon on cross-modal contrastive learning, we decompose the embeddings as
\begin{align}
    t_i &= \mu_t + t'_i \label{eq:mean_decomposition1} \\
    v_i &= \mu_v + v'_i, \label{eq:mean_decomposition2}
\end{align}
where $\mu_v, \mu_t$ denote the modal means of video and text embeddings: 
\begin{align}
    \mu_t &= \frac{1}{B} \sum_{i=1}^B t_i, \\
    \mu_v &= \frac{1}{B} \sum_{i=1}^B v_i,
\end{align}
and $v'_i, t'_i$ represent the displacements from their respective modal means such that $\mathbb{E}[t_i'] = \mathbb{E}[v_i'] = 0$. \Cref{fig:tsne} perceptually illustrates the decomposition.

Using this modal mean decomposition, the cross-modal similarity between a text-video pair can be written as
\begin{align}
    \langle t_i, v_j \rangle = 
    \langle \mu_t, \mu_v \rangle + \langle \mu_v, t_i' \rangle + \langle \mu_t, v_j' \rangle + \langle t_i', v_j' \rangle. 
    \label{eq:logit_decomposition}
\end{align}
The first term is constant irrespective of $i$ or $j$ and can be discarded without loss of generality. The second and third terms are implicit video- and text-wise \textit{biases} that depend either on $i$ or $j$. The last term is the text-video alignment between the displacements from their respective modal means. This decomposition shows that when there are clustering structures within different modalities, there are implicit instance-wise biases that control the overall retrieval weights of each instance. 

For example, we can rewrite the text-to-video retrieval distribution using the decomposition:
\begin{align}
    &P_{t2v}(v_j | t_i) \\
    &= \frac{\exp(\langle t_i, v_j \rangle)^{1/\gamma}}{\sum_{k=1}^B \exp(\langle t_i, v_k \rangle)^{1/\gamma}} \\
    &= \frac{\exp(\langle \mu_t, v'_j \rangle + \langle t'_i, v'_j \rangle)^{1/\gamma}}{\sum_{k=1}^B \exp(\langle \mu_t, v'_k \rangle + \langle t'_i, v'_k \rangle)^{1/\gamma}} \\
    &= \frac{\beta_{j} \exp(\langle t'_i, v'_j \rangle)^{1/\gamma}}{\sum_{k=1}^B \beta_{k} \exp(\langle t'_i, v'_k \rangle)^{1/\gamma}},
    \label{eq:bias_weight}
\end{align}
where $\beta_{j} = \exp(\langle \mu_t, v'_j \rangle)^{1/\gamma}$ and other common terms have been canceled out.
\Cref{eq:bias_weight} can be seen as a weighted Softmax and reveals how the video bias $\langle \mu_t, v'_j \rangle$ controls the weights of the videos in text-to-video retrieval. 

The role of these biases in cross-modal contrastive learning can be better understood through gradient analysis of the cross entropy loss. Denote the video bias as $b_j = \langle \mu_t, v_j' \rangle$. The gradient with respect to the bias is
\begin{align}
    \frac{\partial \mathcal{L}}{\partial b_j} 
    &= -\frac{1}{2B} (1 - \sum_{i=1}^B P_{t2v}(v_j | t_i)), 
    \label{eq:gradient_bias}
\end{align}
where $\sum_{i=1}^B P_{t2v}(v_j | t_i)$ is the retrieval probabilities of video $j$ summed over the set of text queries. \Cref{eq:gradient_bias} shows that the video bias will be learned to minimize the normalization error of the sum retrieval probability. Therefore cross-modal contrastive learning actually learns to normalize the sum retrieval probabilities for each instance during training. However, we empirically find that cross-modal contrastive learning suffers from significant normalization errors of retrieval probabilities and that the explicit normalization proposed in the paper substantially improves its retrieval performance. 
For the reason why cross-modal contrastive learning suffers from such significant normalization errors in practice, we leave it to future work. 

\end{document}